\documentclass[twocolumn,aps,showpacs,floatfix,pra,10pt]{revtex4-1}

\usepackage[latin1]{inputenc}
\usepackage[T1]{fontenc}
\usepackage{amsmath}
\usepackage{amssymb}
\usepackage{latexsym}
\usepackage{ifpdf}
\usepackage{graphicx}

\begin{document}

\title{Casimir torque between nanostructured plates}

\author{R. Gu\'erout}
\affiliation{Laboratoire Kastler Brossel, UPMC-Sorbonne Universit\'es, CNRS, ENS-PSL Research University, Coll\`{e}ge de France,4 Place Jussieu, 75005 Paris, France}
\author{C. Genet}                     
\affiliation{ISIS \& icFRC, University of Strasbourg and CNRS - 8, all\'ee Gaspard Monge, 67000 Strasbourg, France}
\author{A. Lambrecht}
\affiliation{Laboratoire Kastler Brossel, UPMC-Sorbonne Universit\'es, CNRS, ENS-PSL Research University, Coll\`{e}ge de France,4 Place Jussieu, 75005 Paris, France}
\author{S. Reynaud}
\affiliation{Laboratoire Kastler Brossel, UPMC-Sorbonne Universit\'es, CNRS, ENS-PSL Research University, Coll\`{e}ge de France,4 Place Jussieu, 75005 Paris, France}

\begin{abstract}
We investigate in detail the Casimir torque induced by quantum vacuum fluctuations between two nanostructured plates. Our calculations are based on the scattering approach and take into account the coupling between different modes induced by the shape of the surface which are neglected in any sort of proximity approximation or effective medium approach. We then present an experimental setup aiming at measuring this torque. 
\end{abstract}

\pacs{42.50.-p, 03.70.+k, 68.35.Ct}

\maketitle

Quantum vacuum fluctuations produce Casimir forces when scattered by external boundaries. These boundaries may be macroscopic objects such as plates or microspheres, leading to the Casimir effect \cite{Casimir1948} or they may be of microscopic origin such as atoms or molecules experiencing Van der Waals forces. The dynamics of the Casimir effect between macroscopic boundaries is directly related to their geometries, opening a rich physics with a variety of extremely interesting theoretical predictions \cite{BalianAnnPhys1977}. The interplay between geometry and quantum vacuum fluctuations is a topic of high current interest that could lead to new tests of Quantum Electro-Dynamics (QED).
 
Besides fundamental issues, the interplay has also relevant implications in the applied domain. Important technological progress has lately been achieved at sub-micrometer scales. Nanofabrication techniques offer today the possiblity for designing operational micro- and nano-electromechanical systems (MEMS and NEMS) \cite{EkinciRSI2005}. As the dominant interaction between two neutral non-magnetic objects in the sub-mirometer range, the Casimir energy is now recognized as a new mean for actuation, control or functionalization in the actual design of MEMS and NEMS \cite{ChanScience2001}. New possibilities of nano-mechanical control are opened when shaping the interacting objects \cite{CapassoIEEE2007}. 

Due to the dependence of the Casimir energy to boundaries, the source of control is not limited to the normal Casimir effect between planar dielectric or metallic plates but has been extended recently by the observation of the lateral force between corrugated plates \cite{ChenPRL2002,ChiuPRB2009}. This lateral force stems from the translational symmetry breaking along the plates by the imprint of parallel and periodic corrugations on one or both plates. It has been evaluated in the past for perfect boundaries using the path integral approach \cite{EmigPRL2001,EmigPRA2003}  as well as for more realistic dielectric materials  \cite{RodriguesPRL2006,RodriguesPRA2007} within the scattering approach \cite{LambrechtNJP2006, EmigPRL2007}. The latter allows to compute Casimir forces between arbitrarily shaped bodies taking into account the material properties. The lateral force has been shown to be maximized if rectangular structures are imprinted  \cite{LussangePRA2012}. The lateral force has been proposed to be employed in non-contact rack-and-pinion devices with high potential for applications\cite{GolestanianPRL2007,EmigPRL2007b,GolestanianPRE2010}. 

Recently, exact expressions of the Casimir force between dielectric nanostructures have been developed for the normal and the lateral regime \cite{LambrechtPRL2008,DalvitPRA2010,LussangePRA2012,DalvitPRA2012,GueroutPRA2013} that compare very well with recent experiments performed with deep  \cite{ChanPRL2008,ChiuPRB2010} and shallow corrugations \cite{BaoPRL2010}. 

\begin{figure}[htbp]
    \begin{center}
      \includegraphics[width=6cm]{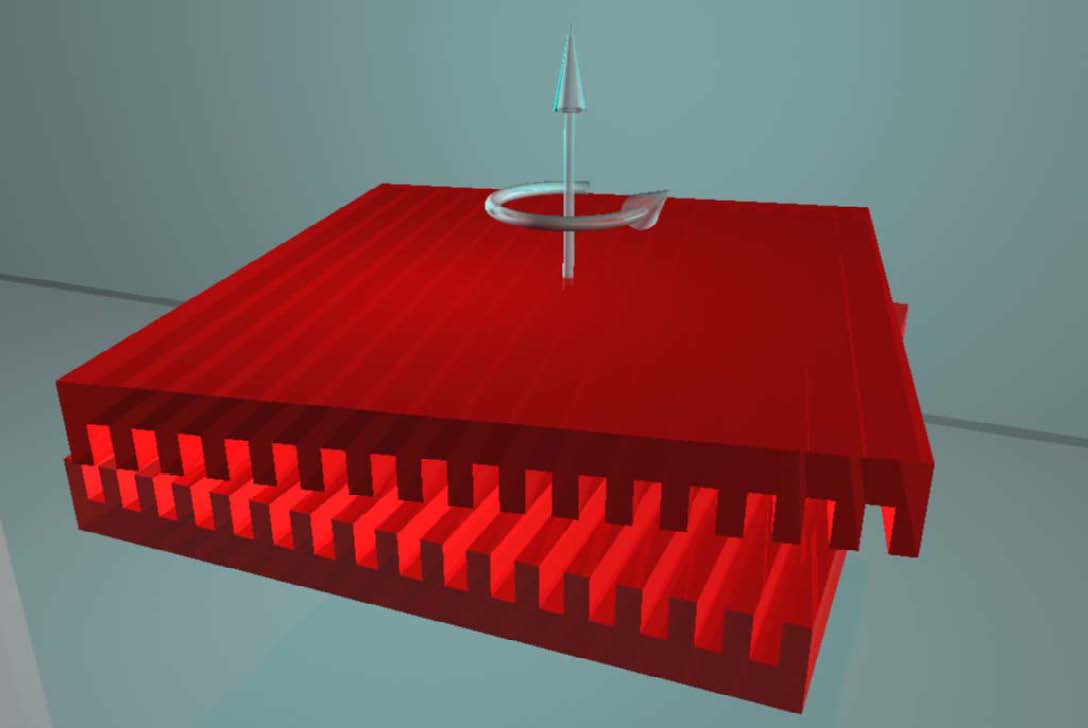}
    \end{center}
    \caption{Sketch of the configuration giving rise to a Casimir torque.}
    \label{fig:T}
  \end{figure}

A different situation of great interest is encountered when the parallel and periodic imprints between two plates are tilted by a small angle, such as shown in Fig. \ref{fig:T}. In this case a vacuum induced Casimir torque arises, which is the main issue of the present paper.   
Interesting perspectives for Casimir torques lie in the design of noncontact gears \cite{MiltonPRD2008a,MiltonPRD2008b} which would allow for torque transmission between two nanostructured plates avoiding any direct contact between them.

As the rotational symmetry is broken in this case, the Casimir torque arises because the Casimir energy now depends on the angle $\theta$ between the two directions of the corrugations. A preliminary study of this situation has already been performed for corrugations small compared to the other length-scales such as the gratings' period, the separation distance or the characteristic wavelength of the material \cite{RodriguesEPL2006} .

Casimir torques have also been studied between anisotropic/birefringent or meta-materials \cite{CapassoPRA2005,CapassoNJP2006,LeonhardtPRA2008,SpencePSSA2013, MundayPRA2015} where the anisotropy or the misalignement between the two optical axes of two plates in vacuum gives rise to a vacuum fluctuation induced torque. The effect is however much smaller and we will discuss the orders of magnitude below in this paper.

Casimir torque measurements between nanostructured plates could constitute an efficient alternative way for testing non-trivial geometrical dependences of the Casimir effect. As the lateral force, the torque has the advantage that it can be measured without any reference to the normal Casimir force. However, if the lateral Casimir force was measured by a clever adaptation of a standard atomic force microscope (AFM) experiment \cite{ChenPRL2002}, torsional motions can not be easily accessed on such systems. The experimental strategy must therefore be completely different from the one adopted for the lateral force. We show later-on that torsion balance techniques are perfectly suited to such experimental aims. 

Theoretical expressions for the Casimir energy $E_{\rm Cas}$ between corrugated plates have been given, based on a general non-specular scattering approach of the Casimir effect \cite{LambrechtNJP2006}. These expressions, specifying the angular mismatch dependence of $E_{\rm Cas}$, have allowed evaluating induced torques up to second order in the corrugation amplitudes \cite{RodriguesEPL2006}. This result contains the pure lateral force as the $\theta=0$ special case and the dependence on the material properties and corrugation period $\lambda_{C}$ of both lateral forces and torques are determined by identical response functions in the perturbative regime \cite{RodriguesPRL2006}. 

Casimir torques have not yet been computed beyond the perturbative regime. But as clearly demonstrated theoretically \cite{GueroutPRA2013} and experimentally \cite{ChanPRL2008,ChiuPRB2010} with normal and lateral Casimir forces, shape effects turn out to be more dramatic when the aspect ratio of the corrugations increases. By analogy, new dynamical regimes are expected at the level of torques that are induced between deep corrugations with respect to torques induced between shallow ones. This could be particularly true when considering metallic plates. It has indeed been realized that the transition from shallow to deep metallic gratings is accompanied by important optical changes \cite{KreiterPRB2002}. Starting with weakly bounded surface plasmon resonances defined on shallow gratings, propagating along the corrugations, new families of resonances are accessible on deep gratings which are related to the localization of the electromagnetic field inside the grooves \cite{BolzhevolnyiPRL2001}. In the limit of one-dimensional slit arrays with sub-micron dimensions, scattering amplitudes are strongly enhanced at resonant wavelengths \cite{GarciaVidalRMP2010}. A Casimir torque will necessarily be modified by the contributions of such surface and guided resonances, suggesting important possibilities for tailoring the dynamical signatures associated to Casimir torques. 

\begin{figure}[htbp]
    \begin{center}
      \includegraphics[width=6cm]{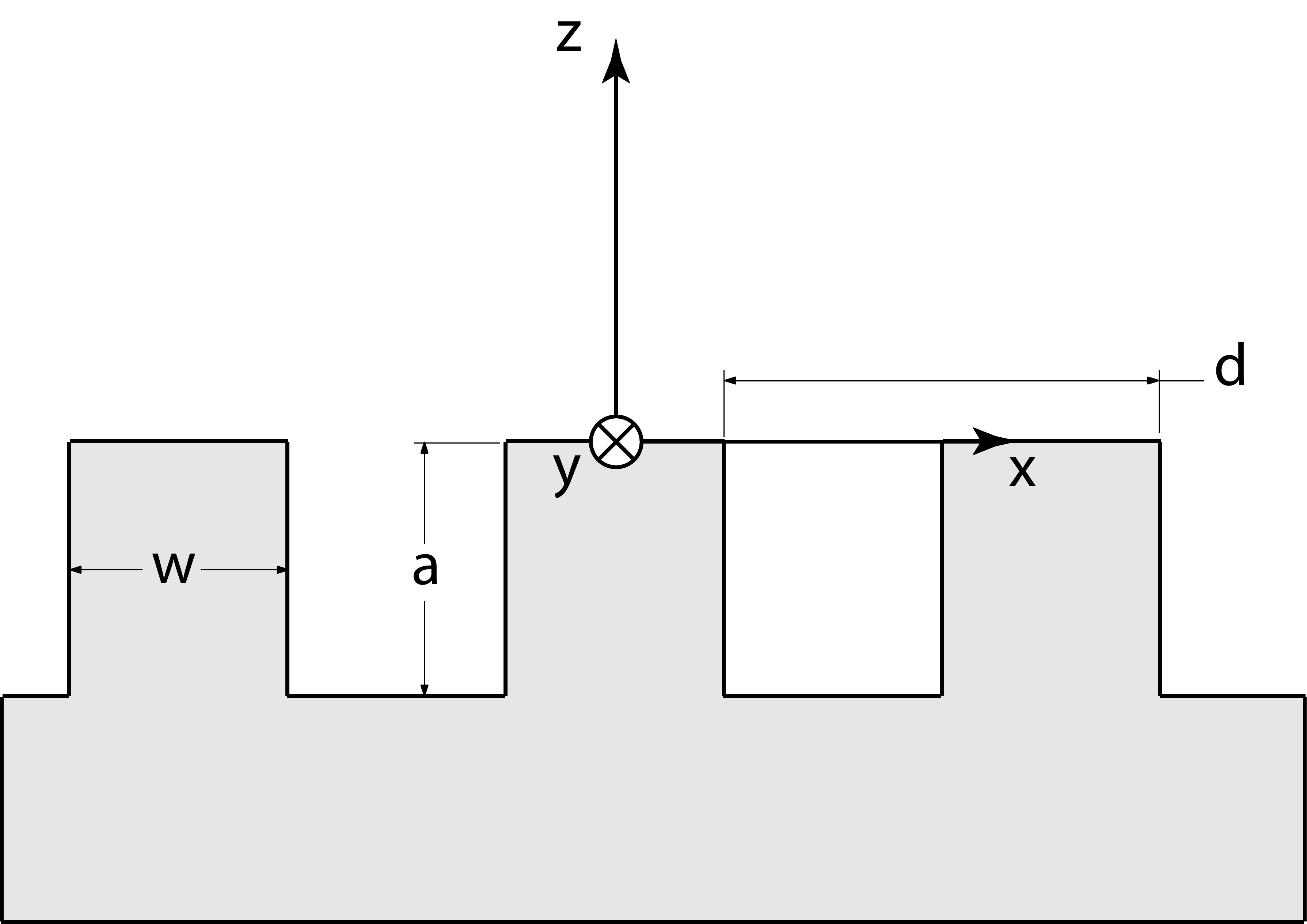}
    \end{center}
    \caption{Schematic representation of a single grating with the geometrical parameters: $a$ and $w$ are respectively the corrugation amplitude and width,  $d$ is the grating's period. }
    \label{fig:SG}
  \end{figure}

We will now present calculations of the Casimir torque between two rectangular gratings. Notations of the parameters are indicated in the schematic view of a single grating on Fig. \ref{fig:SG}. For simplicity the two gratings are supposed to have the same geometric parameters. 

\begin{figure}[htbp]
    \begin{center}
      \includegraphics[width=6cm]{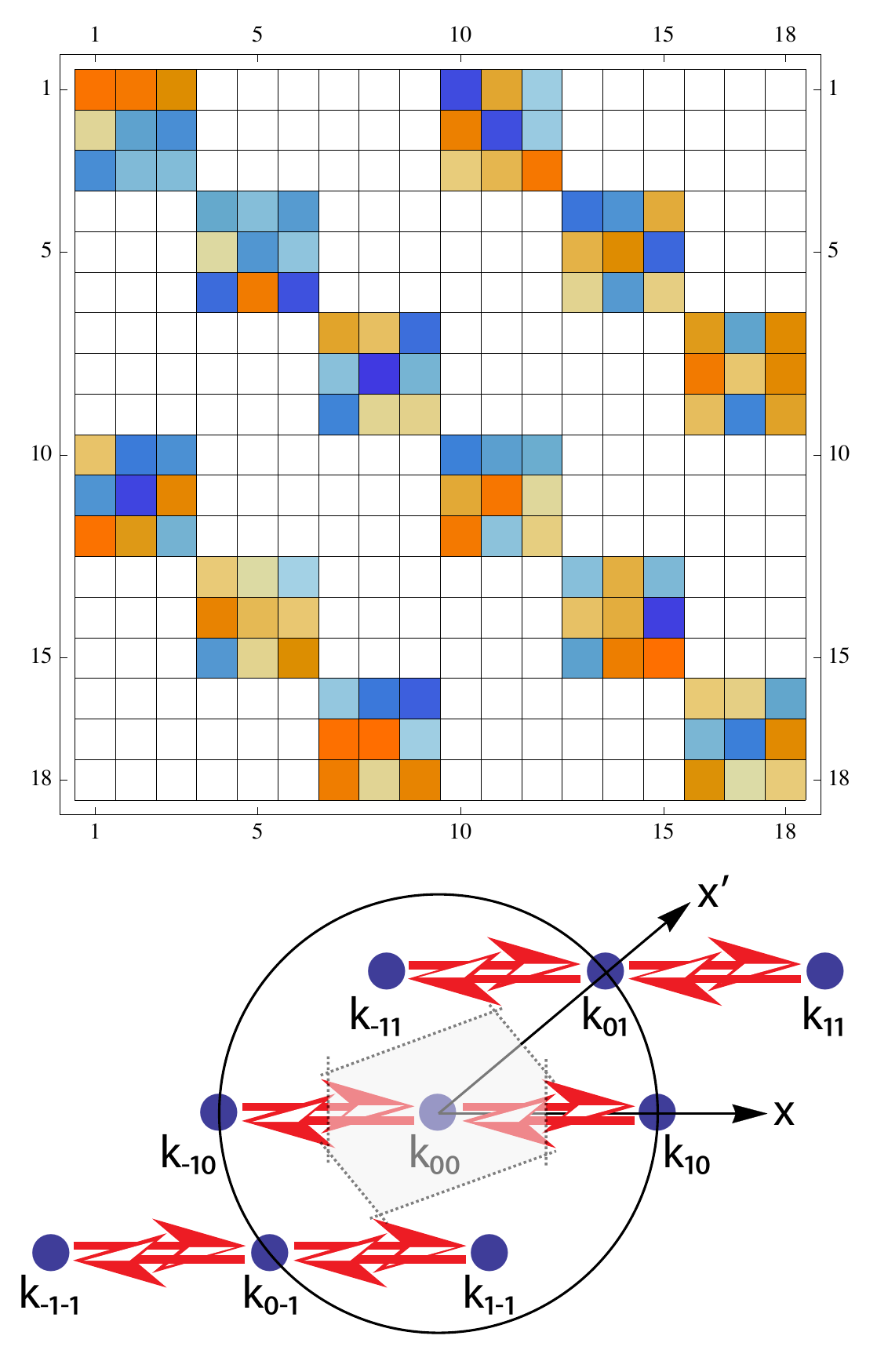}
    \end{center}
    \caption{Schematic structure of the operator $\mathbf{R}_{1}$ in a minimal basis $|n,m,\sigma\rangle$ with $N=1$ whose size is $2(2N+1)^{2}=18$. The zeros are represented in white. The lower part shows the wavevectors which are coupled by diffraction with red arrows. The FBZ is indicated as the grey region.}
    \label{fig:R1}
  \end{figure}

The Casimir torque $\tau$ is defined as  
\begin{equation}
\label{eq:torque}
\tau=-\partial_{\theta}E_{Cas}
\end{equation}

\noindent where $E_{Cas}$ is the Casimir energy and $\theta$ the torsion angle between the two gratings. At zero temperature the Casimir energy between the two corrugated plates reads \cite{GueroutPRA2013}
  \begin{equation}
    \label{eq:CasNRJ}
    E_{Cas}=\frac{\hbar c}{8 \pi^{3}}\int\text{d}\xi\iint_{\text{FBZ}}\text{d}^{2}\mathbf{k}\ln\det\left[\mathbf{1}-\mathbf{R}_{1}\boldsymbol{\mathcal{T}}\mathbf{R}_{2}\boldsymbol{\mathcal{T}}\right].
  \end{equation}
 It is written in terms of the purely imaginary spatial frequency $\imath \xi$ after a Wick rotation. With respect to \cite{GueroutPRA2013} we have performed a change of variable $\xi/c \rightarrow \xi$ here. The operator $\boldsymbol{\mathcal{T}}$ represents a translation of length $L$ from the first to the second grating and $\mathbf{1}$ is the identity operator. $\mathbf{R}_{1}$ and $\mathbf{R}_{2}$ are the reflection operators of the two gratings the derivation of which has been detailed elsewhere~\cite{GueroutPRA2013}. The $\mathbf{k}$-integration runs over the first Brillouin zone (FBZ). The FBZ is a hexagon in the $(k_{x},k_{y})$ reciprocal space as depicted on Fig.~\ref{fig:R1}, which is defined as 

\begin{eqnarray}
\label{FBZ}
&&\{\left(-\frac{\pi}{d}\leq k_{x}\leq \frac{\pi}{d}\right)\cap \\
&&\left(k_{x} \tan\frac{\theta}{2} +\frac{2 \pi}{d}\tan\frac{\theta}{2}\leq k_{y}\leq  k_{x} \tan\frac{\theta}{2} -\frac{2 \pi}{d}\tan\frac{\theta}{2} \right)\cap \nonumber \\
&&\left(-k_{x} \cot \theta +\frac{\pi}{d}\csc \theta \leq k_{y}\leq - k_{x} \cot \theta - \frac{\pi}{d}\csc \theta  \right)\}. \nonumber
\end{eqnarray} 

If we choose the lines of the gratings to be along the $y$ axis, the first grating will couple via diffraction the wavevectors $\mathbf{k}+n\frac{2 \pi}{d}\hat{\mathbf{e}}_{x}$ whereas the rotated grating will couple the wavevectors $\mathbf{k}+n \frac{2 \pi}{d}\hat{\mathbf{e}}'_{x}$ where $\hat{\mathbf{e}}'_{x}=\cos \theta \hat{\mathbf{e}}_{x}+\sin \theta \hat{\mathbf{e}}_{y}$. The index $n$ belongs to the interval $[-N,N]$ with $N$ the highest diffraction order retained in the calculation. 

The common basis between $\mathbf{R}_{1}$ and $\mathbf{R}_{2}$, including two states of orthogonal polarizations $\sigma=\{e,h\}$, is the generalized outer product $(\mathbf{k}+n\frac{2 \pi}{d}\hat{\mathbf{e}}_{x})\otimes(\mathbf{k}+m \frac{2 \pi}{d}\hat{\mathbf{e}}'_{x})\otimes\left(e,h\right)$ of dimension $2(2N+1)^{2}$ which we will refer to as $|n,m,\sigma\rangle$. This is the basis in which the operators appearing in eq.~(\ref{eq:CasNRJ}) are represented. For instance, we therefore have 

\begin{eqnarray}
\label{eq:matrixelements}
\langle n,m,\sigma|\boldsymbol{\mathcal{T}}|n',m',\sigma'\rangle&=&e^{-\kappa_{nm}L}\delta_{nn'}\delta_{mm'}\delta_{\sigma\sigma '}\\
\kappa_{nm}^{2}&=&\xi^{2}+\mathbf{k}_{nm}^{2}\nonumber \\
\mathbf{k}_{nm}&=&\mathbf{k}+\frac{2 \pi}{d}\left(n\hat{\mathbf{e}}_{x}+m\hat{\mathbf{e}}'_{x}\right). \nonumber
 \end{eqnarray}

  \begin{figure}[htbp]
    \begin{center}
      \includegraphics[width=6cm]{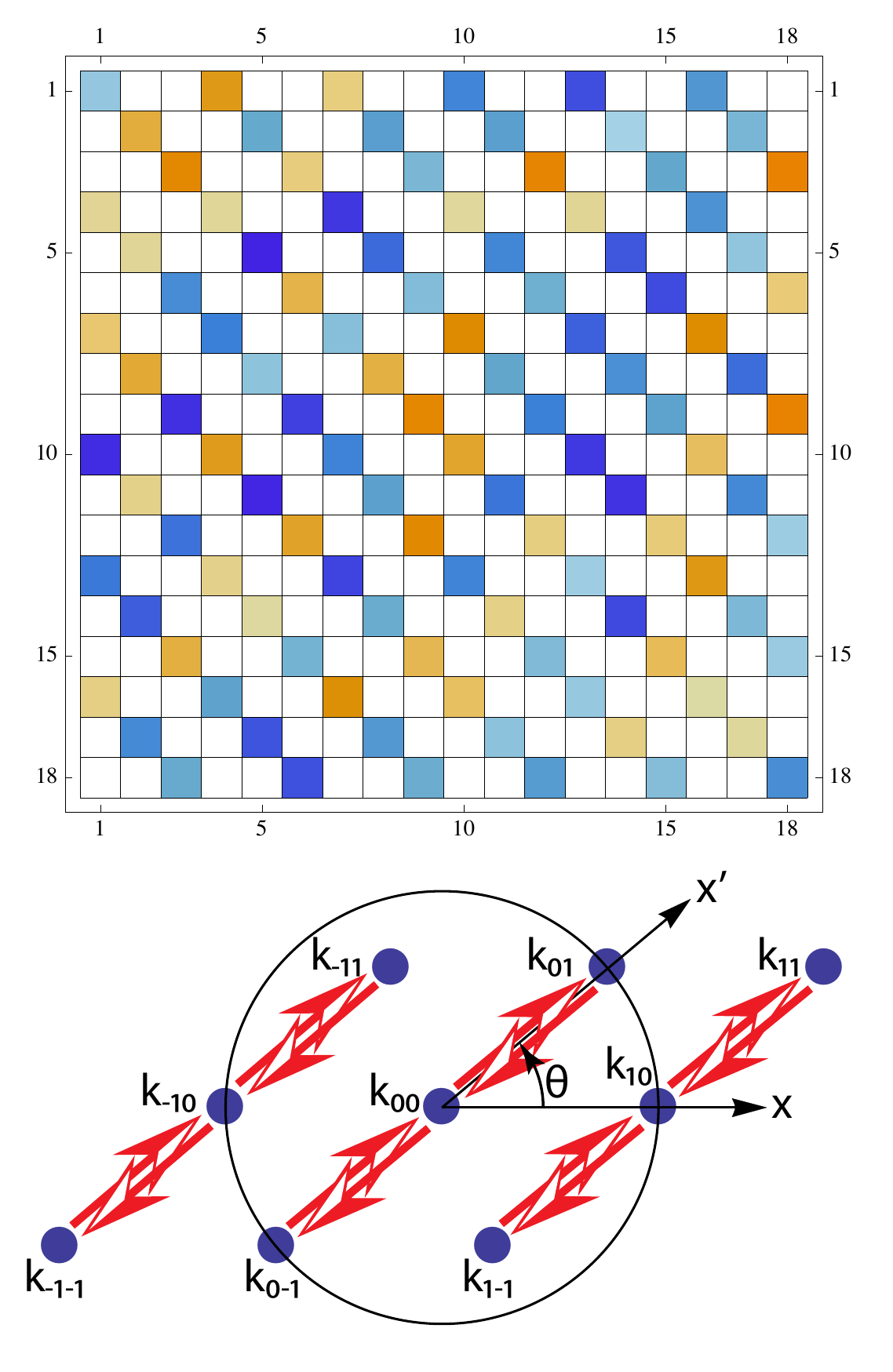}
    \end{center}
    \caption{Same as fig.~\ref{fig:R1} for the operator $\mathbf{R}_{2}$.}
    \label{fig:R2}
  \end{figure}
  
For illustration purposes we show in Figs.~\ref{fig:R1} and~\ref{fig:R2} the schematic structure of the operators $\mathbf{R}_{1}$ and $\mathbf{R}_{2}$ when represented in a minimal basis $|n,m,\sigma\rangle$ for the first diffraction order $N=1$. The first grating represented by the operator $\mathbf{R}_{1}$ diffracts in the direction $\hat{\mathbf{e}}_{x}$ as shown in the lower part of Fig.~\ref{fig:R1}. Therefore $\langle n,m,\sigma|\mathbf{R}_{1}|n',m',\sigma'\rangle\propto \delta_{mm'}$. On the other hand, the rotated grating represented by the operator $\mathbf{R}_{2}$ in fig.~\ref{fig:R2} diffracts only in the direction $\hat{\mathbf{e}}'_{x}$ and we have $\langle n,m,\sigma|\mathbf{R}_{2}|n',m',\sigma'\rangle\propto \delta_{nn'}$. This leads to the structure shown in the upper part of Fig.~\ref{fig:R2}.
  
The area $\mathcal{A}$ of the FBZ is $\mathcal{A}=\frac{4 \pi^{2}}{d^{2}}\sin \theta$. The area covered in $\mathbf{k}$-space by the integration in eq.~(\ref{eq:CasNRJ}) is therefore $(2N+1)^{2}\mathcal{A}$. A prohibitively large number of diffraction order $N$ becomes necessary at a small torsional angle $\theta$ in order to sample the values of the integrand which contribute to the Casimir energy as $\mathcal{A}\to 0$. This stems from the fact that for the combined system of the two rotated gratings, there is a divergent effective period $\frac{d}{\sin \theta}$ in the $y$-direction. For this reason, small torsional angles are numerically unreachable by eq.~(\ref{eq:CasNRJ}). We choose to restrict our calculations for $\theta\geq 5^{\circ}$. Note that this formalism does not gracefully tend towards the formalism for unrotated gratings as $\theta\to 0$ since the FBZ does not tend towards the FBZ$_{0}$ for unrotated gratings which is simply $\{-\frac{\pi}{d}\leq k_{x}\leq \frac{\pi}{d},-\infty\leq k_{y}\leq\infty \}$. This is due to the fact that we consider here gratings with infinite dimensions in the $x$ and $y$ directions.

 We have calculated the Casimir energy $E(\theta)$ as a function of the torsional angle between two gold gratings. The parameters of the gratings are a period $d=400$ nm, a filling factor $f=0.5$ and the height of the corrugations is $a=200$ nm leading to an aspect ratio of 1. The separation distance is $L=100$ nm.
  \begin{figure}[htbp]
    \begin{center}
      \includegraphics[width=8cm]{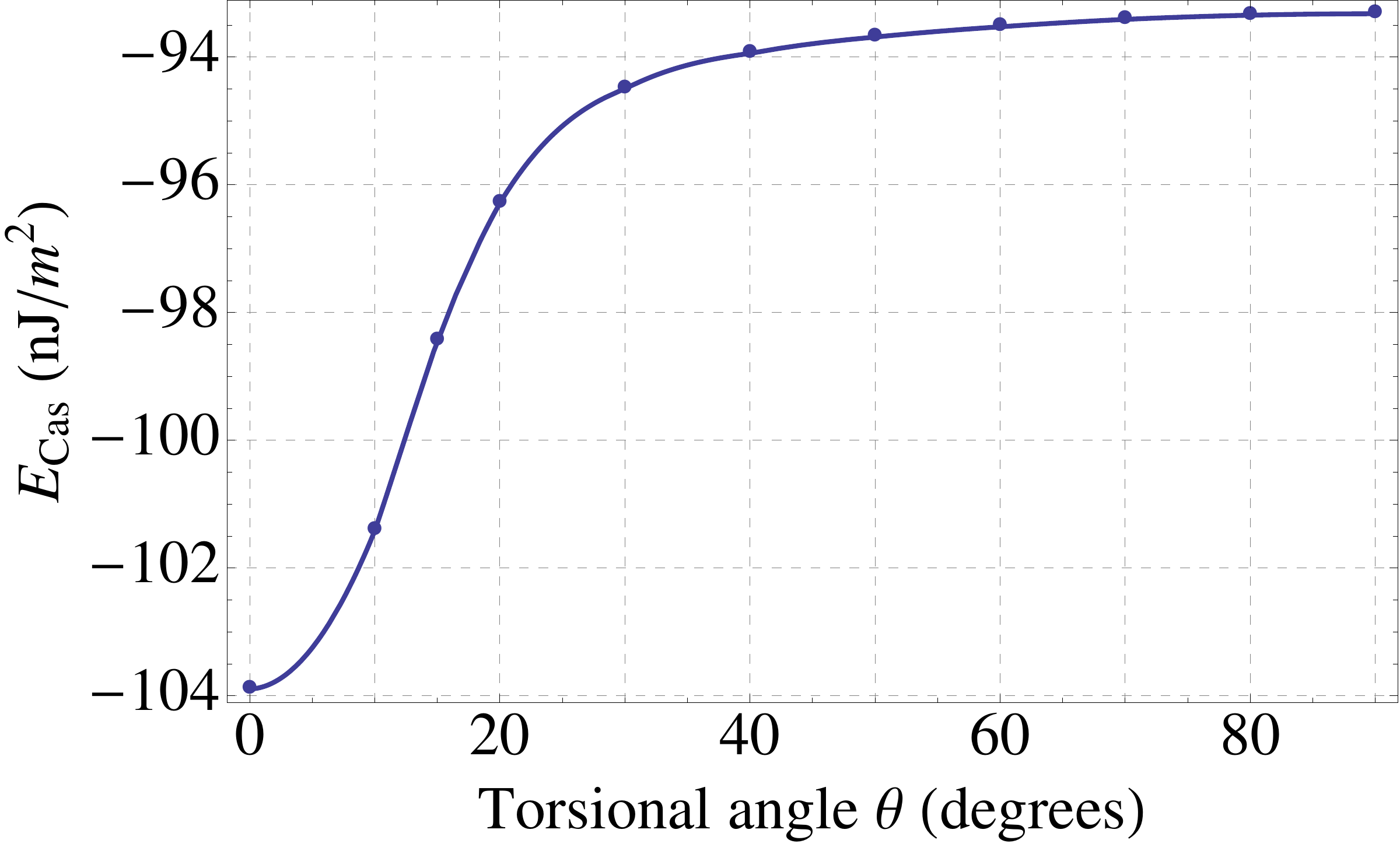}
    \end{center}
    \caption{The Casimir energy $E_{Cas}(\theta)$ as a function of the torsional angle $\theta$ between two gold gratings.}
    \label{fig:couple}
  \end{figure}
 \begin{figure}[htbp]
    \begin{center}
      \includegraphics[width=8cm]{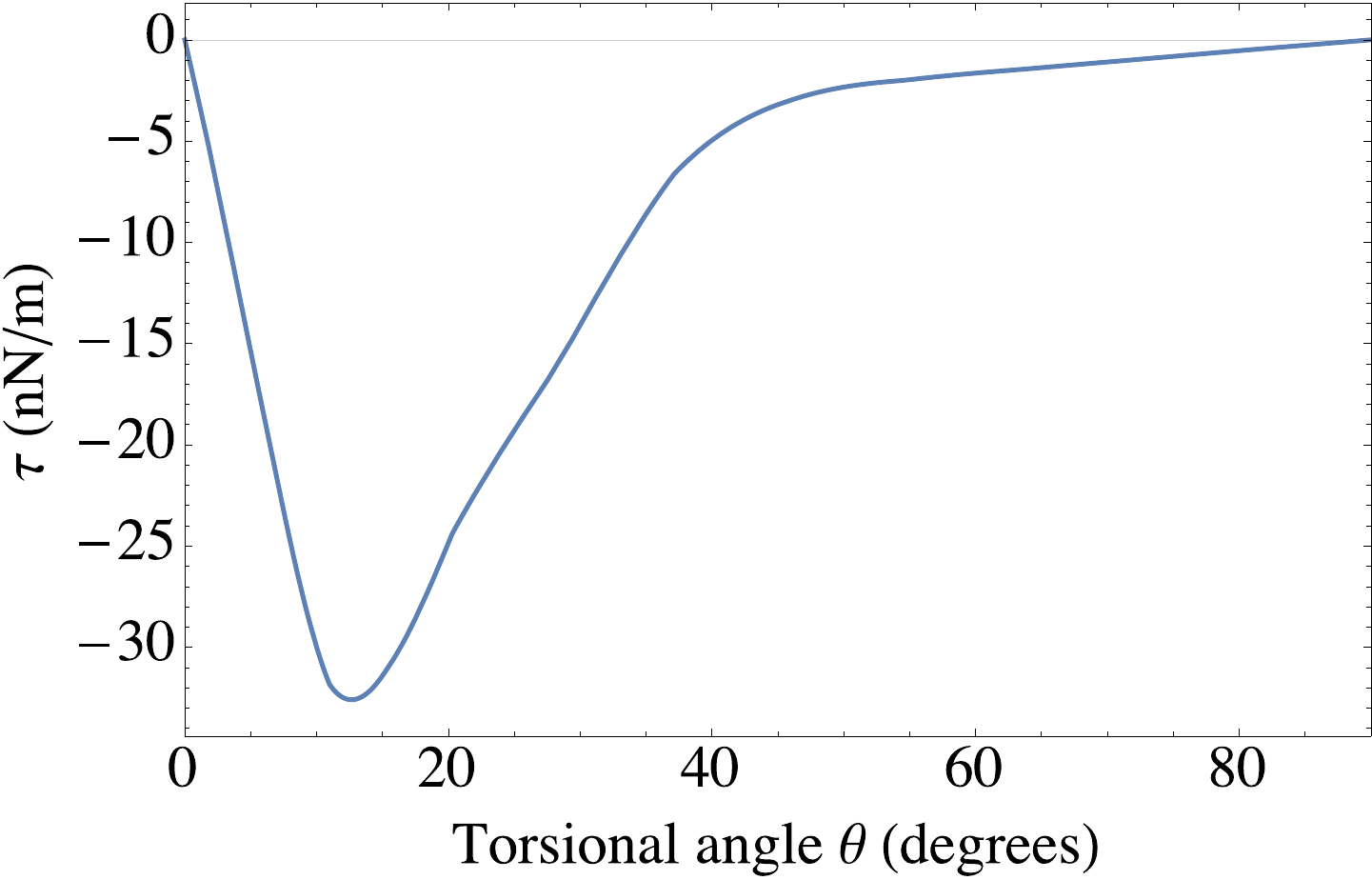}
    \end{center}
    \caption{The Casimir torque $\tau(\theta)$ as a function of the torsional angle $\theta$ between two gold gratings.}
    \label{fig:couple2}
  \end{figure} 
The results are shown in Figs.~(\ref{fig:couple}) and (\ref{fig:couple2}). From these figures, we infer a maximum torque per unit area of $3.5\times 10^{-8}$~N.m$^{-1}$ for a torsional angle $\theta\approx12.5^{\circ}$. Considering a grating surface of ca. 1~mm$^2$, such a level of torque intensity appears well within the range of detectability of the torsion balance setup we present below. This comparably high value should be contrasted with values of Casimir torques induced between anisotropic dielectric plates \cite{CapassoPRA2005,CapassoNJP2006,LeonhardtPRA2008,SpencePSSA2013,MundayPRA2015} which are about 3 orders of magnitude smaller when compared at the same distance. We have also performed calculations for larger corrugations height up to $a=2$~$\mu$m merely leading, somewhat counter-intuitively, to a marginal increase in the Casimir torque.

So far the torque per unit area values reported between birefringent/anistotropic plates are $< 4\times 10^{-10}$~N.m$^{-1}$ at distances of the order of 100~nm and thus turn out to be too faint for an experimental demonstration of this birefringent effect. In a recent paper \cite{MundayPRA2015} an experiment at much shorter distances (10-50~nm) has been proposed, however leading only at the shortest distance of 10 nm to a birefringent torque comparable to the one between nanostructuerd plates at $L=100$~nm. 

\begin{figure}[htbp]
  \begin{center}
    \includegraphics[width=8.5cm]{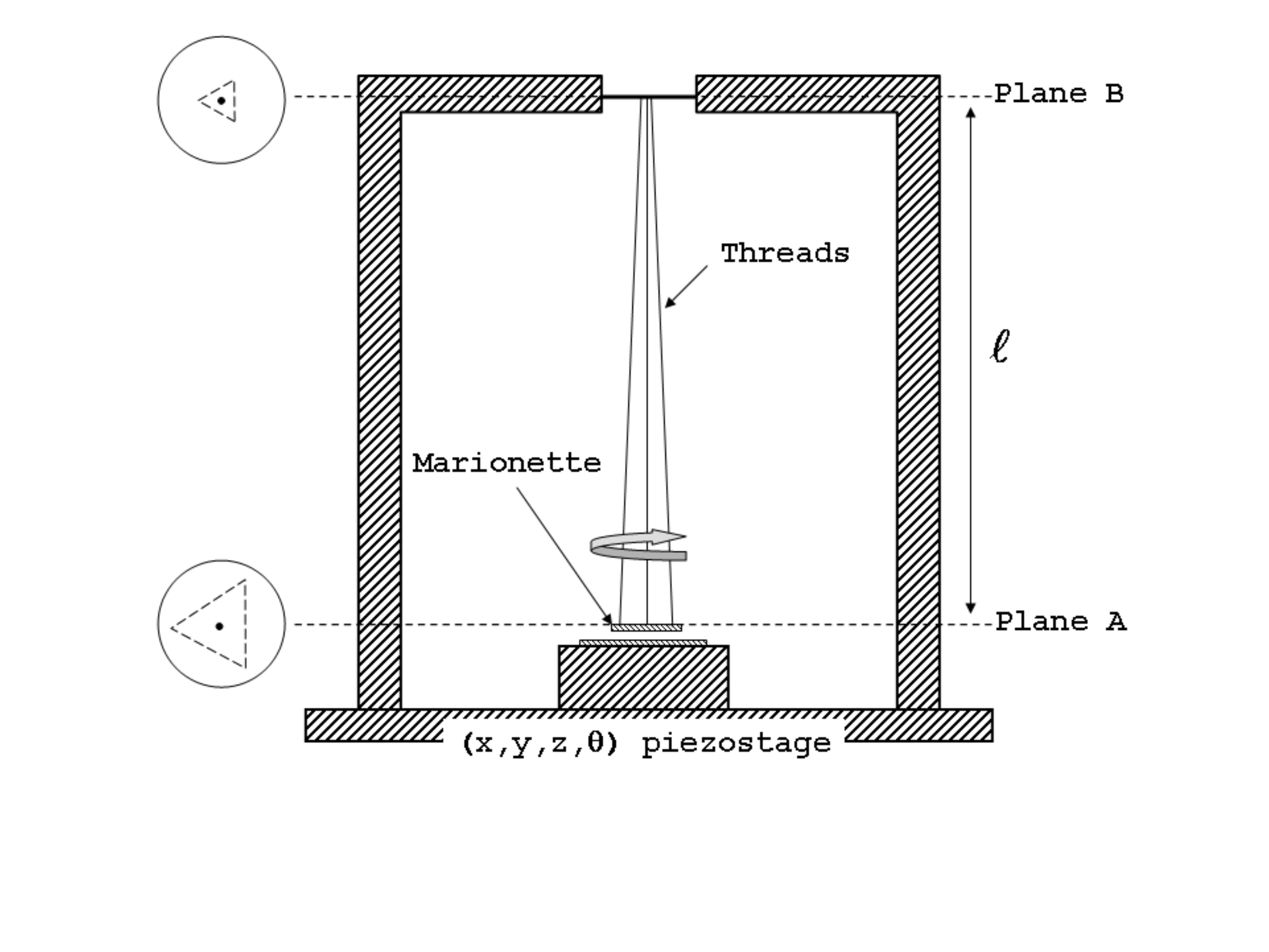}
  \end{center}
  \caption{The marionette is a suspended thin cylinder, of typical mass $0.5$~g and radius $\rho = 5$~mm. The cylinder can be a dielectric (e.g. intrinsic Si) as such, or coated by a thick layer of metal (e.g. Au). One-dimensional corrugations are imprinted on the lower side of the marionette which is then suspended by $\ell=20$~cm from a rigid frame by three threads with a 3-fold rotational symmetry (symbolic triangles on the left-hand side of the frame). On plane A, the supporting points are separated from the symmetry axis of the setup by $a$ and $b$ for the hanging points taken on plane B. Typical values can be chosen as $a=1$~mm and $b=60$~$\mu$m. The substrate is an identical plate with the same corrugations than that of the marionette. Relative spatial and angular positions between the plates are controlled by a high-precision piezo-electric stage. The whole frame rests on an efficient vibration isolation platform and is put in highly quiet environmental conditions.}
  \label{Fig:setup}
\end{figure}

In this context, we propose the design of a torsion balance aimed at measuring weak torques and more specifically Casimir torques between nanostructured plates. Our marionette setup is schematically described in Fig.\ref{Fig:setup}. 

\noindent It consists in a small cylindrical plate, the marionette, suspended above a grounded plate, the substrate, whose relative positioning
is controlled via a piezo-stage. With nanostructures imprinted on their facing sides, the plates will essentially interact through the Casimir energy at sufficiently small gaps. Different types of material can be considered, from intrinsic Si to metal coated surfaces and nanostructures can be milled using for e.g. high resolution focused ion beam (FIB) or electronic beam lithographies.

The marionette in a given plane A is suspended from a fixed frame in plane B by three threads displayed around the axis of the balance with a $2\pi / 3$ rotational symmetry. This tripod geometry is the key feature of our setup since it makes it sensitive to torsional degrees of freedom while being gravity dominated. Indeed, small rotations of our suspended marionette are essentially translated via the tripod geometry into gravity pendulum-like displacements of each thread. As a consequence, the potential of the suspension is determined by the vertical position of the center of mass of the marionette with the important consequence that the intrinsic torsional spring constants of each thread do not enter into the description of the whole suspension setup. This clearly contrasts with classical torsion balance setups where the potential is actually governed by the intrinsic torsional spring constant of the hanging fiber, putting the effort there on seeking for the smallest torsional damping factors and for the largest leverages \cite{LamoreauxPRE2005,LambrechtCQG2005}. In our case, the leverage corresponds to the lateral extensions of the corrugations themselves, lengths that can only be of the order of the mm range. This stresses the crucial need for an original and highly sensitive design.

Yet, as for any suspended system, both rotational and translational degrees of freedom have to be considered here. Induced forces $F$ and torques $\tau$ will be inferred from lateral displacements $\xi$ and deflection angles $\theta$. Corresponding sensitivities will be measured respectively by an effective stiffness $k=F / \xi$ and an effective torsion coefficient $\kappa = \tau / \theta$. In the static case, one has $k=M\omega_{t}^2$ and $\kappa=I\omega_{r}^2$, where $M$ is the mass and $I=M\rho^{2} / 2$ the moment of inertia of the marionette (a cylinder of radius $\rho$), $\omega_{t}$ and $\omega_{r}$ the frequencies of the translation and rotation proper modes. It is important for torque detection to favor sensitivity to rotation over translation degrees of freedom by decoupling them from each other. This is particularly critical in the context of Casimir torques as shown in \cite{RodriguesEPL2006} where the Casimir interaction between two corrugated metallic plates can display a complex surface potential corresponding to possible mixing between rotations and lateral displacements. This has to be controlled in the experiment, and the chosen dimensions of the suspension can already help in this direction. The mass and dimensions $(M,\rho)$ of the marionette, the length $\ell$ of the threads and the positions of the fixation points relative to the rotational axis of the suspension ($a$ and $b$ respectively in plane A and plane B) can indeed be chosen carefully so that $\omega_{t}\gg \omega_{r}$. Calculating the geometry gives $\omega_{r}^2 / \omega_{t}^{2} = 2 a b / \rho^{2}$, with $\omega_{t}^{2}$ simply given by $g / \ell$, $g$ being the acceleration of gravity. The decoupling condition then corresponds to $a,b << \rho$.

When considering realistic values of the parameters, as detailed in the caption of Fig.\ref{Fig:setup}, we reach a ratio between the two frequencies larger than $10$ with $\omega_{r} / 2\pi \sim 0.08$Hz. Having the lowest frequency, the rotation mode is naturally the most easily excited one, and the choice of the $a$ parameter gives an additional freedom to tune $\omega_{r}$ to a specific frequency range. From an experimental point of view, it is particularly interesting to be able to set the rotational frequency to the lower part of the spectrum, below the microseismic fundamental mode at $\sim 0.1$ Hz, hence isolating rotations from most of the seismic background.  

Provided that by design, the rotational degrees of freedom are kept well separated from the translational ones, we model our balance as a damped torsional oscillator with an effective damping factor $\gamma$. The equation of motion of the suspended mirror under the influence of both thermal fluctuations and the applied Casimir torque is thus
  \begin{equation}
	  I\ddot{\theta} + \gamma\dot{\theta} + \kappa\theta = \tau_{\rm th}+\tau_{\rm Cas}.
  \end{equation}

  Moment of inertia $I$ and torsion coefficient $\kappa$ are determined from the setup dimensions and set the rotation proper mode frequency $\omega_{r}$. We can estimate the performances of this ideal system starting from noise limitations in the free pendulum. In this case, with no external torque applied, the system is only driven by thermal fluctuations, and the brownian angular noise spectrum is given as
  \begin{equation}
	  S_{\theta}(\omega)=\frac{2k_{\rm B}T\gamma}{(\kappa-I\omega^{2})^{2}+\gamma^{2}\omega^{2}},  \label{Stheta}
  \end{equation}

 \noindent with $k_{\rm B}$ the Boltzmann constant and $T$ the absolute environmental temperature. The Casimir torque is measured through the angular displacement $\theta$ of the suspended mirror. The estimation of the minimal detectable torque depends on the measurement method and different schemes can be considered \cite{ChenCLASSQUANGRAV1990}. We focus here on measuring the applied torque by detecting quasi-static changes in the angular position of the suspended mirror. The least detectable torque in this situation is given from the low-frequency limit of Eq.(\ref{Stheta}) as \cite{BraginskyBook}
  \begin{equation}
	  \delta\tau_{\rm min}(\omega\ll\omega_{r})\sim \sqrt{2k_{\rm B}T\gamma}  \  \  {\rm N.m} / \sqrt{\rm Hz}
  \end{equation}

 At first sight, the sensitivity of the balance improves as the measurement time is increased and the damping factor reduced. However, achieving better sensitivity by reducing $\gamma$ (i.e. increasing the quality factor of the torsion balance) can turn detrimental at low frequencies since the dynamical response of the now high quality factor oscillator will not be stable over such long measurement times. In this situation, it becomes necessary to increase the bandwidth of the system, through an active control of the damping factor. Such an active control (known as cold-damping technique) increases the damping factor without adding new fluctuations to the system and allows stabilizing the mechanical response of the whole balance. This technique is efficiently implemented in high-precision opto-mechanical \cite{ZendriQG1996,CohadonPRL1999,MarquardtRMP2014} and electro-mechanical experiments \cite{ZolingenPhys1953,ZolingenPhys1953b,TouboulEPJD2000}. 

We can still give simple estimations regarding the potential sensitivity of our setup, assuming, possibly under active control, critical damping with $\gamma / 2= I\omega_{r}$, in which case $\delta\tau_{\rm min}\sim \sqrt{4k_{\rm B}TI\omega_{r}}$ ${\rm N.m} / \sqrt{\rm Hz}$. With the figures given above, this corresponds to $10^{-14}$ ${\rm N.m} / \sqrt{\rm Hz}$ for the least detectable torque. This value nicely falls within the theoretical evaluation of the Casimir torque given above. With actual numbers that will eventually be chosen experimentally in order to maximize the sensitivity (while keeping systematic effects at a minimal level), this preliminary value is already comparable to the sensitivity of existing high precision torsion balances \cite{LambrechtCQG2005,HoylePRD2004}.

An angular change of the suspended mirror can be recorded by an associated displacement of a weak laser beam reflected from the center of the mirror onto a low-noise position sensitive detector. Such a technique can reach a typical sensitivity of a few nrad/$\sqrt{\rm Hz}$ \cite{MuellerOptLett2008}. This detection scheme can be surpassed by optical interferometers \cite{GundlachOL2011}. Interestingly, interferometers can be made sensitive to both rotational and lateral displacements of the marionette. Being additionally well adapted for implementing an active damping of the system, the interferometers can provide a full control over the excitation modes of the balance, as they simultaneously allow for monitoring the displacements of the marionette and for injecting a back-action torque for active control. Also, as well-known when operating MEMS and NEMS and well-recognized in all Casimir force measurements, bolometric and electrostatic backgrounds can easily dominate to screen the dynamical signal \cite{SpeakePRL2003}. They will have to be kept under careful control, though residual spurious effects might easily be rejected from the data using the $\pi$-periodicity of the torque signal \cite{IannuzziSSC2005}. 

With such a high sensitivity, a first Casimir torque demonstration therefore appears within experimental reach. With sufficient confidence in the sensitivity, it might even become possible to work with relatively large plate separations. While this would obviously reduce the torque intensity, it would have important experimental advantages, gaining over two major technical issues: a large working distance reduces tilt effects on the torque due to deviation from strict parallelism and minimizes the problem of interstitial dust particles, too. Finally, we note that here we have limited the discussion to quasi-static measurements. One has to keep in mind that dynamic mode detection can potentially yield better sensitivity, in particular as far as $1/ f$ noise limitation is concerned.

\end{document}